\documentclass[a4paper,11pt]{article}
\usepackage{setspace}
\usepackage{jcappub} 
\usepackage[export]{adjustbox} 
\usepackage[utf8]{inputenc}
\usepackage{color}
\usepackage{graphicx}
\usepackage{amsmath,amsfonts,amssymb}
\usepackage{acronym}
\usepackage{xspace}
\usepackage{multirow}
\usepackage{tabularx}
\usepackage{orcidlink}
\usepackage{ragged2e}
\usepackage{enumitem}
\usepackage{caption}
\usepackage{subcaption}
\usepackage{booktabs}
\usepackage{comment}

\usepackage{listings}
\usepackage[ruled,vlined]{algorithm2e}

\SetCommentSty{mycommfont}

\hypersetup{
     colorlinks   = true,
     citecolor    = violet,
     urlcolor     = violet,
     linkcolor    = violet}

\newcommand{\msun}{M_{\odot}}

\newcommand{\dd}{\mathrm{d}}

\def\lsim{\mathrel{\rlap{\lower4pt\hbox{\hskip0.5pt$\sim$}}
 \raise1pt\hbox{$<$}}}         
\def\gsim{\mathrel{\rlap{\lower4pt\hbox{\hskip0.5pt$\sim$}}
 \raise1pt\hbox{$>$}}}         

 \def\be   {\begin{equation}}   \def\ee   {\end{equation}}
 \def\ba   {\begin{array}}      \def\ea   {\end{array}}
 \def\bea  {\begin{equation}\begin{aligned}}   \def\eea  {\end{aligned}\end{equation}}
 \def\bean {\begin{eqnarray*}}  \def\eean {\end{eqnarray*}}

\definecolor{TardisBlueAcceso}{RGB}{0, 80, 200}

\definecolor{oucrimsonred}{rgb}{0.6, 0.0, 0.0}
\definecolor{persianblue}{rgb}{0.11, 0.22, 0.73}
\definecolor{forestgreen}{rgb}{0.13,0.35,0.13}
\definecolor{lightgray}{rgb}{0.83, 0.83, 0.83}
\definecolor{oucrimsonred}{rgb}{0.6, 0.0, 0.0}

\newcommand{\Msun}{M_\odot}
\newcommand{\Ogw}{\Omega_{\rm GW}}
\newcommand{\rISCO}{r_{\rm ISCO}}
\newcommand{\nISCO}{f_{\rm ISCO}}
\newcommand{\rISSO}{r_{\rm ISSO}}

\title{All Dark Matter as Primordial Black Holes: a Gravitational-Wave Test}

\author[1]{A. Ianniccari\orcidlink{0009-0008-9885-7737}}
\author[2]{A.J. Iovino\orcidlink{0000-0002-8531-5962}}
\author[3]{G. Perna\orcidlink{0000-0002-7364-1904}} 
\author[4]{D. Perrone\orcidlink{0000-0003-4430-4914}}
\author[1]{A. Riotto\orcidlink{0000-0001-6948-0856}}

\affiliation[1]{Department of Theoretical Physics and Gravitational Wave Science Center, 24 quai E. Ansermet, CH-1211 Geneva 4, Switzerland}
\affiliation[2]{New York University, Abu Dhabi, PO Box 129188 Saadiyat Island, Abu Dhabi, UAE}
\affiliation[3]{Keemilise ja Bioloogilise F\"u\"usika Instituut, R\"avala pst. 10, 10143 Tallinn, Estonia}
\affiliation[4]{Center of Gravity, Niels Bohr Institute, Blegdamsvej 17, 2100 Copenhagen, Denmark}

\abstract{We propose a novel  test for the hypothesis that   the dark matter of the Universe is composed  entirely by asteroidal-mass primordial black holes. It relies on detecting   the stochastic gravitational-wave background generated by the coalescence onto   rapidly spinning supermassive astrophysical black holes of the   light primordial black holes    
accumulated in   spherically symmetric spikes. We discuss the scenarios in which the signal can  be  observed by the future space-based gravitational-wave observatories  LISA  and DECIGO.}

\begin{document}
\maketitle
\flushbottom

\section{Introduction}\label{sec:intro}
The idea that black holes may have formed in the early Universe instead of as the final state of stellar evolution, goes back to the late sixties and early seventies~\cite{Zeldovich:1967lct,Hawking:1971ei,Carr:1974nx,Carr:1975qj}, and the suggestion that such primordial black holes (PBHs) could account for the dark matter (DM) without invoking any new elementary particle followed almost immediately~\cite{Chapline:1975ojl,Meszaros:1974tb} (for recent reviews on the topic see Refs.~\cite{Carr:2016drx,Carr:2020xqk,Carr:2026hot,Riotto:2024ayo}).

The formation channels that have been proposed are numerous and predict very different mass spectra. The most studied one is the direct collapse of a large curvature perturbation generated during inflation once the corresponding scale re-enters the cosmological horizon, which, to allow a copious production of PBHs, requires an enhancement of the power spectrum with respect to the amplitude measured on cosmic microwave background scales. In practice, this can be realised either through a transient violation of the slow roll dynamics as the inflaton field evolves through its potential~\cite{Ivanov:1994pa} or by the presence of a spectator field~\cite{Lyth:2001nq,Enqvist:2001zp,Moroi:2001ct,Lyth:2002my,Wands:2002bn,Gordon:2002gv,Enqvist:2003mr,Dimopoulos:2003ss,Dimopoulos:2003az,Bartolo:2003jx}.

Beyond the collapse of adiabatic perturbations, PBHs may be produced at a first-order phase transitions~\cite{Sato:1981gv,Kodama:1982sf,Hawking:1982ga,Jedamzik:1996mr,Jedamzik:1999am}, from the collapse of cosmic string loops~\cite{Hawking:1987bn,Polnarev:1988dh,Caldwell:1995fu,MacGibbon:1997pu} or of domain wall networks~\cite{Gouttenoire:2023gbn,Ferreira:2024eru,Dunsky:2024zdo}, or during an early matter-dominated or slow-reheating era, where the absence of pressure support makes the collapse far easier and the resulting holes are born with a sizeable spin~\cite{Green:2000he,Harada:2016mhb,Harada:2017fjm,Kokubu:2018fxy,Padilla:2021zgm,DeLuca:2021pls,Padilla:2024iyr}.

As of today, independently of the formation channel, the PBHs are allowed to contribute to the totality of dark matter only in the so-called asteroidal mass window 
\begin{equation}
  10^{-16}\,\Msun\;\lesssim\;m\;\lesssim\;10^{-11}\,\Msun 
  \label{eq:window}.
\end{equation}
Indeed, below roughly $10^{-16}\,\msun$ the Hawking evaporation time drops below the age of the Universe, so that the PBHs have either disappeared already or are still radiating enough to leave an imprint on the Galactic and extragalactic photon backgrounds~\cite{Auffinger:2022khh}. Above roughly $10^{-11}\,\Msun$, instead, the Einstein radius of the lens becomes large compared with the projected radius of the source star, finite-source effects no longer wash out the magnification, and the absence of microlensing events towards the Magellanic Clouds and M31 closes the window from the opposite side~\cite{Niikura:2017zjd,Mroz:2024mse}.

However, the production of Gravitational Waves (GWs) could be a complementary tool to explore this mass range. In particular if the DM is composed solely of PBHs, then they would accumulate in the center of the galaxies, around astrophysical supermassive black holes (SMBHs)~\cite{Xu:1994vb,Carr:1997cn,Gondolo:1999ef,Carr:2020gox} and fall inside it, in a coalescence process~\cite{Eda:2013gg,Kuhnel:2018mlr}.
This scenario is interesting since the relevant orbital frequencies are of the order of the millihertz, probed by LISA~\cite{LISA:2017pwj}, and the mechanism that brings PBHs into their immediate neighbourhood is already part of the standard astrophysical dynamics.

A SMBH that grows adiabatically drags the surrounding DM inwards and leaves behind a steep density spike, whose slope and inner boundary have been computed in full general relativity for both the Schwarzschild and the Kerr geometry~\cite{Gondolo:1999ef,Sadeghian:2013laa,Ferrer:2017xwm}. 
If the DM is made of PBHs, the spike is a granular population of compact objects on bound orbits, and those that end up close enough to the SMBH lose energy to GWs, spiralling inwards until they cross the innermost stable orbit and are swallowed~\cite{Nishikawa:2017chy,Feng:2024obn}. Each of these systems is an extreme mass-ratio inspiral with $q=m/M\sim10^{-22}\text{--}10^{-17}$, radiating in the millihertz band because the orbital frequency is fixed by $M$ alone. However, the single signals are far too weak to be resolved individually under any assumption, so that the real observable is the incoherent superposition of signals emitted by the entire population over all the galaxies in the Universe, in the same sense in which the stellar-mass objects inhabiting the same nuclei produce an astrophysical EMRI background~\cite{Bonetti:2020jku,Pozzoli:2023kxy,Caputo:2026rle}.

The aim of this work is to compute the GW background generated by DM PBHs inspiralling into SMBHs and to determine for which SMBH masses and spins, PBH masses, and spike slopes the signal can be detected by LISA~\cite{LISA:2017pwj,LISA:2024hlh} and DECIGO~\cite{Seto_2001, Kawamura:2020pcg}.

This manuscript is organised as follows. In Sec.~\ref{sec:spike} we define the DM spike and solve for its evolution under radiation reaction. In Sec.~\ref{sec:GW_num} we compute the GW background for generic inclined orbits. In Sec.~\ref{sec:equatorial} we focus on the equatorial case, where everything can be done analytically. In Sec.~\ref{sec:results} we discuss the detectability with LISA, and we conclude in Sec.~\ref{Sec::Conclusions}. Some technicalities are present in three appendices.

\section{The DM spike and its evolution}
\label{sec:spike}
 
\subsection{Initial condition}
\label{sec:spike_ic}
 
We assume a spherically symmetric spike composed of PBHs with a power-law profile,
\begin{equation}
  \rho_{\rm PBH}(r) = \rho_{0}\left(\frac{r}{r_{0}}\right)^{-\gamma},
  \qquad r_{\rm in}\le r\le r_{0},
  \label{eq:spike}
\end{equation}
with $\rho_0$ the density at the outer radius $r_0$ and the slope in the range $1\leq\gamma<3$, depending on the astrophysical processes that shape the spike~\cite{Gondolo:1999ef,Ullio:2001fb,Merritt:2002vj,Sadeghian:2013laa,Ferrer:2017xwm,Feng:2024obn}. 

The lower limit $\gamma=1$ corresponds to the inner cusp of a Navarro-Frenk-White halo in the absence of a central SMBH, while $\gamma\to3$ is the upper limit for a power-law spike of finite mass. Two-body relaxation of an equal-mass population of compact objects leads to the Bahcall-Wolf profile, with $\gamma=7/4$~\cite{Bahcall:1976aa,Lightman:1977zz}. The adiabatic growth of a SMBH steepens instead a pre-existing DM cusp into a spike with $2.25\leq\gamma\leq2.5$, depending on the slope of the initial halo~\cite{Quinlan:1994ed,Gondolo:1999ef,Nishikawa:2017chy,Feng:2024obn}\footnote{In strongly mass-segregated DM cases where a light and an heavy component exists, the heavy and rare objects can form even steeper cusps, up to $\gamma\simeq2.75$~\cite{Alexander:2008tq}. In the case in ehich the DM is not only composed of PBHs this scenario could be considered further.}. 

We assume that the PBHs in the spike can be treated as collisionless particles moving along spherical geodesics around central SMBH, to simplify our calculations. 
Therefore we can set the inner boundary $r_{\rm in}$ as the innermost stable orbit.
For equatorial orbits this is the usual innermost stable circular orbit (ISCO), while for inclined orbits it becomes the innermost stable spherical orbit (ISSO), which depends on the inclination angle $\iota$ and lies between the prograde and the retrograde equatorial values,
\begin{equation}
    \rISCO^{\,\rm pro}(1,\chi)\;\le\;\rISSO(\cos\iota ,\,\chi)\;\le\;\rISCO^{\,\rm retro}(-1,\chi),
    \label{eq:isso_range}
\end{equation}
being $\chi$ the dimensionless spin.
The orientation of the orbit is parametrised by $x \equiv \cos\iota$,  where $x=1$ and $x=-1$ correspond to equatorial prograde and retrograde orbits, respectively, and $x$ is a conserved quantity for the timelike geodesics, as explained in Sec. \ref{sec:GW_num}.

 For an isotropic distribution, the probability density in $x$ is uniform,
 \begin{equation}
     P(x)=\frac{1}{2}\;, \qquad \int_{-1}^{1}P(x)\;dx= 1.
 \end{equation}
 For a monochromatic mass function
 \footnote{The monochromatic case allows a simple  mathematical description and we  verified that the impact of an extended mass function does not alter our results.  For a lognormal mass function with unit width, the change would be of about a factor 1.5 and only in the IR tail. This can be understood considering that broadening the mass function would generate PBHs of different masses around the central one which would constitute anyway the majority of the signals. This would not change the peak frequency, which is established by the central SMBH, but can slightly change the number of sources contributing to each frequency bin and the amplitude of the signal they generate, which goes like $\sim M m_{\rm PBH}$.} 
 the number density is $n_{\rm PBH}=\rho_{\rm PBH}/m$, and the initial distribution of the PBHs in a single spike reads
\begin{equation}
  \left.\frac{\dd^2 N}{\dd r\dd x}\right|_{0}
  =\frac{4\pi r^{2}}{m}\rho_{\rm{PBH}}(r)P(x)
  =\frac{2\pi\rho_{0}r_{0}^{\gamma}}{m}\;r^{\,2-\gamma} ,
  \label{eq:dNdr0}
\end{equation}
where the factor $4\pi r^2$ follows from the spherical symmetry of the spike and the factor $P(x)=1/2$ from the isotropy of the orbital orientations. 
We define the total mass of the spike as a fraction $\Delta$ of the SMBH mass, $\Delta\equiv M_{\rm sp}/M$, with
\begin{align}
    M_{\rm sp}
    &=
    m
    \int_{-1}^{1}\dd x
    \int_{r_{\rm ISSO}(\chi,x)}^{r_0}\dd r\,
    \left.\frac{\dd^2N}{\dd r\,\dd x}\right|_0
    \nonumber\\
    &=
    4\pi\rho_0 r_0^\gamma
    \int_{-1}^{1}\dd x\,P(x)
    \int_{r_{\rm ISSO}(\chi,x)}^{r_0}
    \dd r\,r^{2-\gamma}
    \nonumber\\
    &=
    \frac{4\pi\rho_0 r_0^\gamma}{3-\gamma}
    \left[
        r_0^{3-\gamma}
        -
        \int_{-1}^{1}\dd x\,
        P(x)\,
        r_{\rm ISSO}^{\,3-\gamma}(\chi,x)
    \right].
\end{align}
Throughout this work we fix the outer radius of the spike to $r_0=2GM/3\sigma_{\rm DM}^2$, using the relation between the SMBH mass $M$ and the fitted velocity dispersion $\sigma_{\rm DM}$ from Ref.~\cite{Gultekin:2009qn}. 
For $\gamma<3$ the term $r_{\rm ISSO}(\chi,x)$ is always negligible with respect to the outer edge $r_0$, so we can invert invert the relation above to get
\begin{equation}
    \rho_{0}\simeq\frac{\Delta M(3-\gamma)}{4\pi r_{0}^{3}}.
\end{equation}
The spike fraction $\Delta$ is not a free parameter of the model, since a spike massive enough to affect the stellar or gas kinematics would already have been absorbed into the dynamical mass inferred for the central black hole~\cite{Gorchtein:2010xa,Lacroix:2016qpq}. Direct measurements are available only for a  few galaxies, for example $\Delta\lesssim0.5$ for Centaurus\,A~\cite{Cappellari:2008db}, $\Delta\lesssim0.08$ for Sgr\,A$^{*}$~\cite{Gillessen:2008qv} and $\Delta\lesssim0.1$ for M87~\cite{EventHorizonTelescope:2019ggy}, so it can be taken of the order of $0.1$. 
An alternative definition, discussed in Ref.~\cite{Feng:2024obn}, relates $\Delta$ to the slope of the spike through the influence radius of the SMBH~\cite{Merritt:2003qc}. 
The DM mass enclosed within a radius $R$ is
\begin{equation}
    M_{\rm DM}(<R)=4\pi\int^{R}\dd r\;r^2\rho_0\left(\frac{r}{r_0}\right)^{-\gamma}= \frac{4\pi\rho_0r_0^{\gamma}R^{3-\gamma}}{3-\gamma}.
\end{equation}
Defining the influence radius $r_h$ as the radius for which the DM mass is 2 times the central mass object~\cite{Merritt:2003qc} we have
\begin{equation}
    M_{\rm{DM}}(r<r_h)= 2M,
\end{equation}
and we set $r_0= 0.2 r_h$ to close the relation, as in~\cite{Lenoci:2023gjz}.
Extending the profile to $R=r_h$ we obtain
\begin{equation}
    2M=\frac{4\pi\rho_0r_0^{\gamma}}{3-\gamma}\left(\frac{r_0}{0.2}\right)^{3-\gamma}= M_{\rm{sp}}\cdot 0.2^{-(3-\gamma)},
\end{equation}
that leads to
\begin{equation}
    \Delta(\gamma)= 2\cdot 0.2^{3-\gamma}.
    \label{eq:Delta_gamma}
\end{equation}
In the following, we use this relation to compute the amplitude of the signal.

\begin{figure}[htbp]
    \centering
\includegraphics[width=0.8\linewidth]{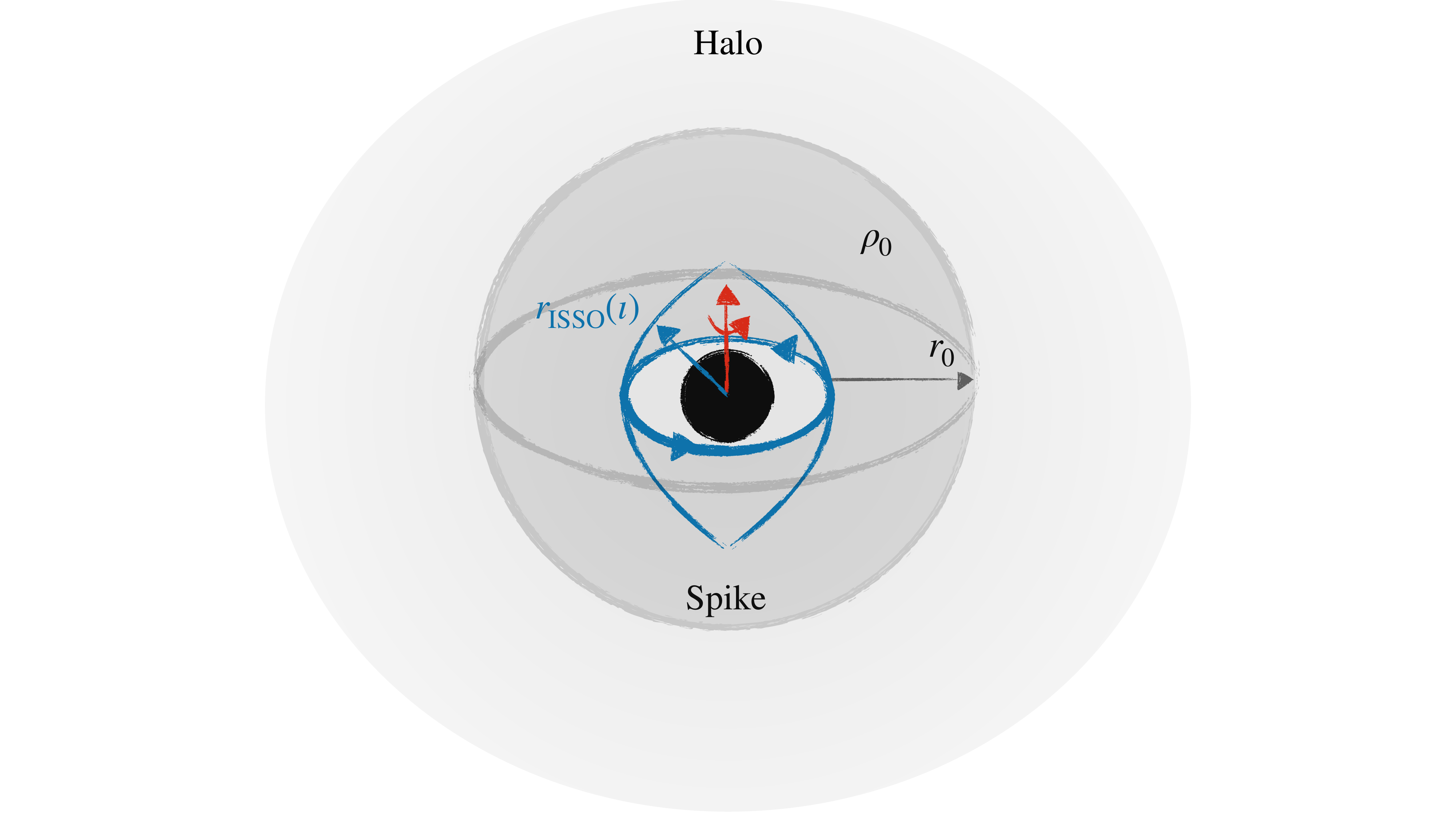}
    \caption{Schematic representation of the density of the spike around a SMBH.}
    \label{fig:placeholder}
\end{figure}

\subsection{Radiation reaction and the temporal evolution of the spike}
\label{sec:drain}
 
For a quasi-circular orbit the radiation reaction integrated between $r$ and the ISSO yields the inspiral time~\cite{Peters:1963ux,Peters:1964zz}
\begin{equation}
  \tau(r)=k\left(r^{4}-\rISSO^{4}(\chi\, ,x)\right),
  \quad \text{with} \quad
  k=\frac{5c^{5}}{256\,G^{3}Mm(M+m)}\;\simeq\;\frac{5c^{5}}{256\,G^{3}M^{2}m}\,,
  \label{eq:taur}
\end{equation}
where in the last step we used $m\ll M$. 
Inverting this relation at fixed inclination gives the radius within which all the PBHs have reached the ISSO, and have therefore merged by the time $\tau(z)$,
\begin{equation}
  r(\tau)=\left(\rISSO^{4}(\chi\, ,x)+\frac{\tau(z)}{k}\right)^{1/4},
  \label{eq:roftau}
\end{equation}
where 
\begin{equation}
  \tau(z)=\int_{z}^{z_{\rm form}}\frac{\dd z'}{(1+z')H(z')} \; 
  \label{eq:rdrain}
\end{equation}
is the cosmic time elapsed since the formation of the spike, which we set at the  fiducial redshift $z_{\rm form}=5$ within the range of formation epochs considered in previous studies of cosmological DM spikes \cite{Belikov_2014, Ahn_2007}. 
In the range $0\leq z\leq z_{\rm form}$ we can write the elapsed time analytically as
\begin{equation}
 \tau(z)
=
\frac{2}{3H_0\sqrt{\Omega_\Lambda}}
\left[
\operatorname{arcsinh}\!\left(
\sqrt{\frac{\Omega_\Lambda}{\Omega_{\rm matter}}}\,(1+z)^{-3/2}
\right)
-
\operatorname{arcsinh}\!\left(
\sqrt{\frac{\Omega_\Lambda}{\Omega_{\rm matter}}}\,(1+z_{\rm{form}})^{-3/2}
\right)
\right],
\end{equation}
where $H_0$ is the Hubble parameter and $\Omega_{\Lambda}, \;\Omega_{\rm matter}$ are the density fractions of the cosmological constant and matter respectively.

We now study how the spike evolves with time. 
A PBH released at radius $r_{i}$ is found at time $\tau$ at the radius $r$ fixed by $\tau=k\,(r_{i}^{4}-r^{4})$, that is
\begin{equation}
  r_{i}(r,\tau)=\left(r^{4}+\frac{\tau}{k}\right)^{1/4},
  \qquad
  \frac{\dd r_{i}}{\dd r}=\frac{r^{3}}{r_{i}^{3}} .
  \label{eq:charflow}
\end{equation}
Since the PBHs are neither created nor destroyed before reaching the inner boundary, their number in corresponding shells is conserved\footnote{We are assuming the orbital inclination to be constant in the in-falling process. In the LISA frequency band, this hypothesis is completely fulfilled, since the spike is frozen for the considered scenario. 
For smaller masses of the central object we keep this simplifying assumption even though there are corrections given by the orbital variation of the trajectory from the emission of gravitational waves.},
\begin{equation}
  \frac{\dd^2 N}{\dd r\,\dd x}(r,x,\tau)\dd r\,\dd x
  =\frac{\dd^2 N}{\dd r_i\,\dd x}(r_i,x)\dd r_i\,\dd x .
\end{equation}
Using the initial isotropic distribution
\begin{equation}
  \left.\frac{\dd^2 N}{\dd r_i\,\dd x}\right|_{0}
  =\frac{2\pi\rho_{0}r_{0}^{\gamma}}{m}\;r_i^{\,2-\gamma},
\end{equation}
we obtain
\begin{equation}
    \frac{\dd^2 N}{\dd r\,\dd x}(r,x,\tau)=\left.\frac{\dd^2 N}{\dd r_i\,\dd x}\right|_{0}(r,x,\tau)\frac{\dd r_i}{\dd r}= \frac{2\pi\rho_0r_0^{\gamma}}{m}r^3\left(r^4+\frac{\tau(z)}{k}\right)^{-(1+\gamma)/4}
    \label{eq:dNdr_evolved}
\end{equation}
on $r_{\rm ISSO}\leq r_i(r,\tau)\leq r_0(M)$. 
For $r^{4}\gg\tau/k$ the spike can be considered frozen\footnote{The evolution of the spike can be neglected for $M\in[5\times10^5,10^7]\,M_{\odot}$, where the LISA sensitivity curve reaches its minimum. Since the inspiral time scales as $M^2/m$, the drain becomes more important for smaller SMBH masses, around $M\simeq 10^4\,M_{\odot}$, which emit in an higher frequency range. For these masses the effect is important and cannot be neglected.}, and Eq.~\eqref{eq:dNdr_evolved} reduces to the initial profile \eqref{eq:dNdr0}.

\section{The GW background from PBH spikes}
\label{sec:GW_num}
\subsection{Spherical orbits in Kerr geometry}
We study now the orbits in a spinning BH background.
The Kerr geometry can be written in Boyer-Lindquist coordinates as
\begin{align}
    \dd s^2 =& - \left(1- \frac{2M r}{\rho^2}\right)\dd t^2 - \left(\frac{4 a M r\sin^2\theta}{\rho^2}\right)\dd t \dd \phi + \frac{\rho^2}{\Delta_{\rm K}}\dd r^2+  \nonumber\\
    &\rho^2 \dd \theta^2 +\sin^2\theta \left(r^2+a^2 + \frac{2 M a^2 r\sin^2 \theta}{\rho^2}\right)\dd\phi^2,
\end{align}
where $M$ is the mass of the black hole, $aM$ is its angular momentum, and
\begin{equation}
\label{eq:kerrfunctions}
    \rho^2=r^2+a^2 \cos^2\theta,\quad \Delta=r^2 -2Mr+a^2.
\end{equation}
A timelike geodesic is characterized by three constants of motion: the specific energy $E$, the specific axial angular momentum $L_{z}$ and the Carter constant $Q$~\cite{Carter:1968rr}. Therefore we define
\begin{equation}
    x \equiv \cos\iota\equiv\frac{L_z}{\sqrt{L_z^2+Q}} \qquad -1\leq x\leq 1 .
\end{equation}
In terms of the Mino time $\lambda$, the radial and polar motions decouple completely,
\begin{align}
\left(\frac{\dd r}{\dd\lambda}\right)^{2}
&= \left[E(r^{2}+a^{2})-aL_z\right]^{2}
-\Delta_{\rm K}\left[r^{2}+(L_z-aE)^{2}+Q\right]
\;\equiv\; R(r), \label{eq:geo_r}\\[4pt]
\left(\frac{\dd\theta}{\dd\lambda}\right)^{2}
&= Q-\cot^{2}\theta\,L_z^{2}
-a^{2}\cos^{2}\theta\left(1-E^{2}\right)
\;\equiv\; \Theta(\theta), \label{eq:geo_theta}\\[4pt]
\frac{\dd\phi}{\dd\lambda}
&= \csc^{2}\theta\,L_z
+aE\left(\frac{r^{2}+a^{2}}{\Delta_{\rm K}}-1\right)
-\frac{a^{2}L_z}{\Delta_{\rm K}}, \label{eq:geo_phi}\\[4pt]
\frac{\dd t}{\dd\lambda}
&= E\left[\frac{(r^{2}+a^{2})^{2}}{\Delta_{\rm K}}
-a^{2}\sin^{2}\theta\right]
+aL_z\left(1-\frac{r^{2}+a^{2}}{\Delta_{\rm K}}\right). \label{eq:geo_t}
\end{align}
where $\dd /\dd\lambda= \Sigma\,\dd /\dd\tau $, with $\tau$ the proper time of the particle.

Orbits with zero eccentricity are spherical, which means that the radius is constant while the particle oscillates in the polar direction between two turning points and rotates in the azimuthal one. 
For a given radius and inclination these orbits are obtained by imposing $R(r)=R'(r)=0$, which fixes the energy $E$ and the axial angular momentum $L_{z}$ \cite{Hughes:1999bq}, while the additional condition $R''(r)=0$ defines the innermost stable spherical orbit $r_{\rm ISSO}(\chi,x)$.
From the equations above we can extract the physical orbital frequencies $\dd \theta/\dd t=\Omega_{\theta}(r,x)$ and $\dd \phi/\dd t=\Omega_{\phi}(r,x)$ for every value of the inclination $x$ and of the spin $\chi$.
We compute these frequencies and $r_{\rm ISSO}(\chi,x)$ with the \texttt{KerrGeodesics} package of the \texttt{Black Hole Perturbation Toolkit}~\cite{Fujita:2009bp,Stein:2019buj}.

\subsection{Energy flux to infinity}
\label{sec:teukolsky_modes}

The GW flux emitted by a point particle moving on a bound Kerr geodesic is computed by solving the Teukolsky equation with a source for the perturbations of the Kerr spacetime~\cite{Teukolsky:1973ha,Sasaki:2003xr,Barack:2018yvs}. 
Since the source motion is quasi-periodic, each mode can be labeled by the quantities $(s,\ell,\tilde m,k,n)$, with $s=-2$ for gravitational perturbations, $\ell$ and $\tilde m$ the spheroidal-harmonic indices and $k$ and $n$ the polar and radial harmonic numbers (we denote the azimuthal number by $\tilde m$ to avoid confusion with the PBH mass $m$), and has frequency
\begin{equation}
f_{\tilde m kn}
=
\tilde m\Omega_\phi
+
k\Omega_\theta
+
n\Omega_r .
\label{eq:teukolsky_frequency}
\end{equation}
We evaluate the modes with the \texttt{TeukolskyPointParticleMode} function of the \texttt{Black Hole Perturbation Toolkit}, keeping only
\begin{equation}
(s,\ell,\tilde m,k,n)
=
(-2,\ell,\pm\ell,0,0)
\label{eq:modes}
\end{equation}
where $n=0$ because the orbits are spherical and $k=0$ is a conservative choice\footnote{A complete calculation would sum over all the modes $(\ell,\tilde m,k)$. For nearly equatorial orbits the emission is dominated by $\tilde m=\pm\ell$, while for polar orbits, $x\simeq0$, the leading contribution to the quadrupole comes from $\tilde m=0$, $k=\pm2$. Given that the largest part of the signal comes from the equatorial prograde orbit we can ignore this additional factor and get a good approximation of the signal as shown in \ref{sec:equatorial}.
We estimate this additional contribution in App.~\ref{app:polar_modes} and include it in the uncertainty on the signal discussed in Sec.~\ref{sec:results}.}. The frequency of the selected modes is
\begin{equation}
f_{\ell}(r,x)
=
\frac{c^3}{2\pi GM}
\left|\ell\;\Omega_\phi(r,x)\right| ,
\label{eq:inclined_frequency}
\end{equation}
where we restored the dimensions. Summing the two contributions $\tilde m=\pm\ell$, the luminosity at infinity of each multipole is
\begin{equation}
L_{\ell}^\infty(r,x)
=
\frac{c^5}{G}
\left(\frac{m}{M}\right)^2
\left[\mathcal F^\infty_{\ell,\ell,0,0}(r,x)+\mathcal F^\infty_{\ell,-\ell,0,0}(r,x)\right]\,.
\label{eq:physical_teukolsky_luminosity}
\end{equation}

\subsection{The inclination-averaged GW background}
\label{sec:inclined_background}

At fixed $x$ the relation \eqref{eq:inclined_frequency} is monotonic and can be inverted numerically to give $r_{\ell}=r(f_{\ell},x)$, so that both the luminosity and the distribution of the PBHs can be written as functions of the frequency. 
The spectral luminosity of a single spike then follows by weighting the emission of each orbit with the number of PBHs that occupy it, summing over the multipoles and integrating over the orientations,
\begin{align}
\mathcal L_f^{\rm sp}[f,z,M]
={}&
\int_{-1}^{1}\dd x\;\sum_{\ell}
L_{\ell}^\infty\left[r(f_{\ell},x),x\right]
\left.\frac{\dd^2N}{\dd r\,\dd x}\right|_{r(f_\ell,x)}
\left|\frac{\partial r}{\partial f_\ell}\right|_x
\nonumber\\
&\times
\Theta\left(f-f_{\rm min}(x,\ell)\right)
\Theta\left(f_{\rm ISSO}(x,\ell)-f\right).
\label{eq:inclination_averaged_luminosity}
\end{align}
where the two Heaviside functions restrict the emission of each multipole to the frequencies spanned, at the given inclination, by the orbits between $r_0$ (which corresponds to $f_{\rm min}$) and the ISSO. 
The background observed today follows from~\cite{Phinney:2001di}, with the source frame frequency redshifted as $f_s=(1+z)f$,
\begin{equation}
\Ogw(f)
=
\frac{f}{\rho_{\rm c}}
\int_0^{z_{\rm form}}\dd z\;\int\dd\ln M\,
\frac{\dd n}{\dd\ln M}\frac{\;\mathcal L_f^{\rm sp}\left[(1+z)f,M\right]}{(1+z)H(z)} 
\label{eq:inclined_omega_gw}
\end{equation}
where $\rho_{\rm c}=3H_0^2c^2/(8\pi G)$ is the critical energy density and $\dd n/\dd\ln M$ the comoving SMBH mass function. 
For the latter we adopt the optimistic scenario of Ref.~\cite{Caputo:2026rle}, based on the occupation-corrected local black-hole mass function of Ref.~\cite{Burke:2024wcf}, and we discuss the impact of this choice in Sec.~\ref{sec:results}.

\section{The equatorial limit}
\label{sec:equatorial}

The equatorial prograde orbit, $x=1$, has the smallest ISCO radius and is therefore the most energetic orbital configuration, providing an upper estimate of the signal. The equatorial retrograde orbit, $x=-1$, has instead the largest ISCO radius and provides a natural lower bound.

Circular equatorial geodesics of the Kerr metric have angular frequency 
\begin{equation}
  {\Omega_\phi}(r,\chi)=\frac{1}{(r/M)^{3/2}+\chi},
  \label{eq:nukerr}
\end{equation}
so that the frequency of the dominant quadrupolar emission is 
\begin{equation}
\label{eq:rad_freq}
    Mf_{s}=\frac{{\Omega_\phi}}{\pi}=\frac{1}{\pi \left[(r/M)^{3/2}+\chi\right]}
\end{equation}
with the convention that $\chi>0$ and $\chi<0$ describe prograde and retrograde orbits around a black hole of spin $|\chi|$.
Inverting the relation between frequency and radius gives
\begin{equation}
  r(f_{s},\chi)=M
  \left(\frac{1}{\pi Mf_{s}}-\chi\right)^{2/3},
  \qquad
  \left|\frac{\dd r}{\dd f_{s}}\right|
  =\frac{2}{3\pi f_{s}^{2}}
   \left(\frac{1}{\pi Mf_{s}}-\chi\right)^{-1/3}.
  \label{eq:rofnu}
\end{equation}
The ISCO radius is given by~\cite{Bardeen:1972fi}
\begin{equation}
  \frac{r_{\rm ISCO}(\chi)}{M}=3+Z_{2}-{\rm sgn}(\chi)\sqrt{(3-Z_{1})(3+Z_{1}+2Z_{2})},
\end{equation}
with
\begin{equation}
  Z_{1}=1+\left(1-\chi^{2}\right)^{1/3}\left[(1+\chi)^{1/3}+(1-\chi)^{1/3}\right] \quad \textrm{and}
  \quad
  Z_{2}=\sqrt{3\chi^{2}+Z_{1}^{2}}.
  \label{eq:Risco}
\end{equation}
The emission extends from the frequency of the outermost orbits, at $r_0$, up to the ISCO frequency,
\begin{equation}
    f_{0}\le f\le f_{\rm{ISCO}},
\end{equation}
where
\begin{equation}
  f_{0}(\chi,M)=\frac{1}{\pi M\left((r_{0}/M)^{3/2}+\chi\right)},
  \qquad
  \nISCO(\chi,M)=\frac{1}{\pi M\left((r_{\rm ISCO}/M)^{3/2}+\chi\right)}.
  \label{eq:cutoffs}
\end{equation}
Every PBH is a source radiating at the monochromatic frequency \eqref{eq:rad_freq}, and the spectral luminosity of the spike around each SMBH is
\begin{align}
  \mathcal{L}_{f}(f_s,z)\simeq\,\,P(x=1)\cdot\frac{\dd N}{\dd r}(r,z)L_{\rm GW}(r)\,\,\left|\frac{\dd r}{\dd f_s}\right|_{\,r=r(f_s,\chi)}, \label{eq:calL}
\end{align}
where we retained only the equatorial prograde orbits, $x=1$, which give the largest contribution, and where we remind that $f_s=(1+z)f$ and $P(x)=\frac{1}{2}$.
We approximate the luminosity with the Newtonian quadrupole formula for circular orbits for simplicity, so we use
\begin{equation}
    L_{\rm GW}(r)=L_{\rm GW}^{\rm Newt}(r)=\frac{32}{5}\,\frac{m^{2}M^{3}}{r^{5}}.
    \label{eq:LNewt}
\end{equation}
Including the evolution of the spike, described in Eq.~\eqref{eq:dNdr_evolved}
\begin{equation}
    \frac{\dd N}{\dd r}(r,z)= \frac{2\pi\rho_0r_0^{\gamma}}{m}r^3\left(r^4+\frac{\tau(z)}{k}\right)^{-(1+\gamma)/4},
\end{equation}
we obtain a closed expression for the spectral luminosity
\begin{align}
   \mathcal{L}_{f}\left(f_s,z,M\right)
=& \frac{128}{15}\frac{G\,m \,c^2}{f_{s}^2}\;
   \rho_{0}\left(\frac{r_{0}}{r_g}\right)^{\gamma}
   \left(\frac{c}{\pi\, r_g f_{s}}-\chi\right)^{-\frac{5}{3}}\left[\left(\frac{r(f_{s},\chi)}{r_g}\right)^{4}+\frac{256}{5}\frac{m}{M}\frac{c \tau(z) }{r_g}\right]^{-\frac{1+\gamma}{4}}
   \nonumber\\
& {}\times
\Theta\!\left[\nISCO-f_s\right]
\Theta\!\left[f_s-f_{s,0}\right].\nonumber\\
&
\end{align}
where we introduced $r_g=GM/c^2$ and restored $G$ and $c$. 
The background then follows from Eq.~\eqref{eq:inclined_omega_gw}.
This approximation allows us to study how the peak of the signal depends on the masses of the SMBH and the PBH and the parameters of the spike. 
We highlight the linear scaling with the PBH mass $m$ in the scenario when the contribution of $\tau(z)$ is negligible, i.e., in the frozen spike limit.
Furthermore the factor $c/\pi r_gf_s-\chi$ shows that at higher spins the signal is boosted, given that the ISCO lies closer to the SMBH horizon.

\begin{figure}[hbt!]
	\begin{center}
\includegraphics[width=.78\textwidth]{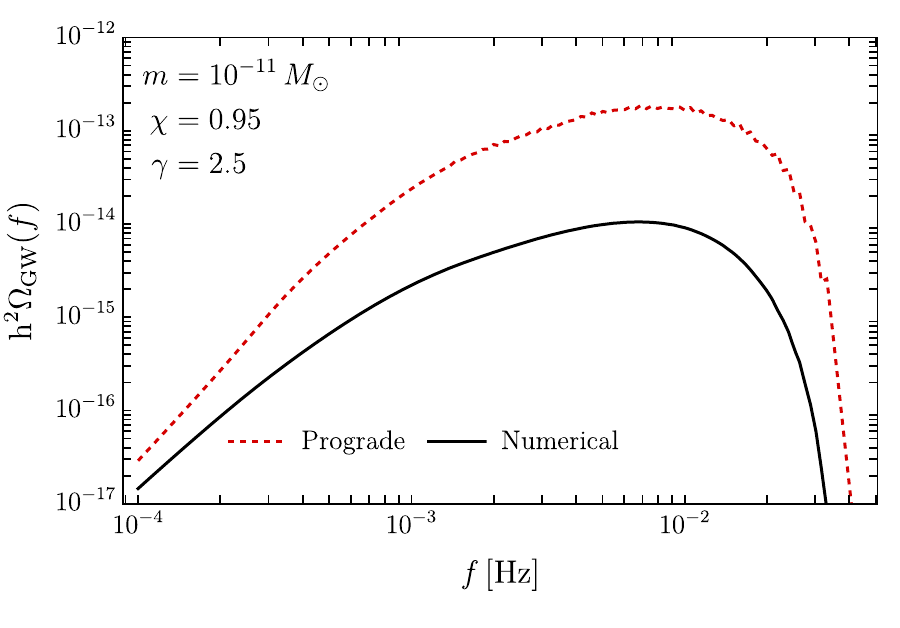}
		\caption{Stochastic GW background generated by the PBH spike for $m=10^{-11}\,M_\odot$, $\chi=0.95$ and $\gamma=2.5$. The dashed red curve is the analytical result for equatorial prograde orbits with the Newtonian luminosity and the solid black curve is the numerical calculation of Sec.~\ref{sec:inclined_background}, where we have considered only the quadrupolar contributions $\ell=2$, $\tilde m=\pm2$.}
        \label{fig:Num}
	\end{center}
\end{figure}

Figure~\ref{fig:Num} compares the GW background obtained from the full numerical calculation of Eq.~\eqref{eq:inclined_omega_gw}, which includes all the orbital inclinations and uses only the quadrupolar modes $\ell=2$, $\tilde m=\pm2$, with the analytical approximation based only on equatorial prograde orbits. 
The analytical result captures the shape of the spectrum quite well but overestimates the peak by about one order of magnitude and shifts it slightly to higher frequencies. 
The difference is due to the Newtonian quadrupole luminosity, which is not reliable in the strong-field region, and we study this correction in App.~\ref{app:flux}.

\section{Prospects for future GW interferometers}
\label{sec:results}

\begin{figure}[htb!]
	\begin{center}
\includegraphics[width=.7\textwidth]{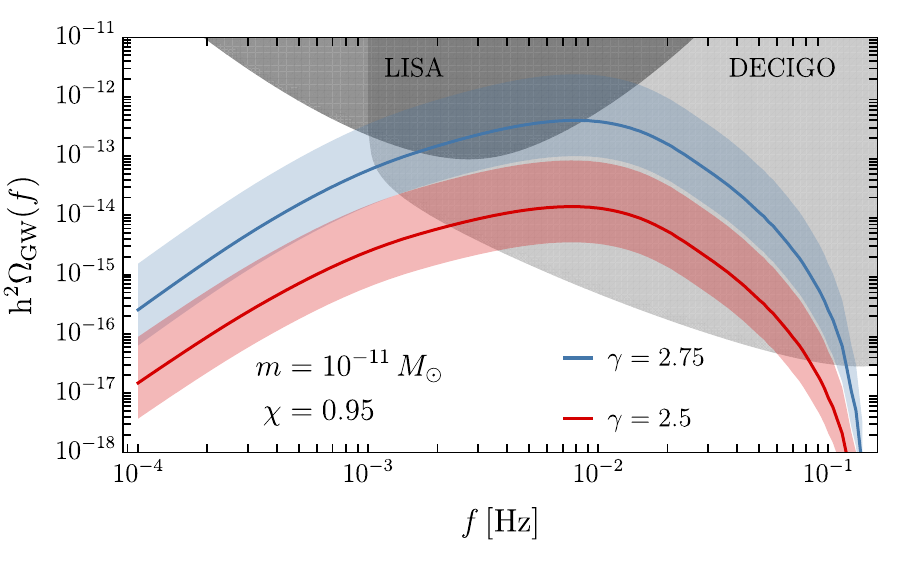}
\includegraphics[width=.7\textwidth]{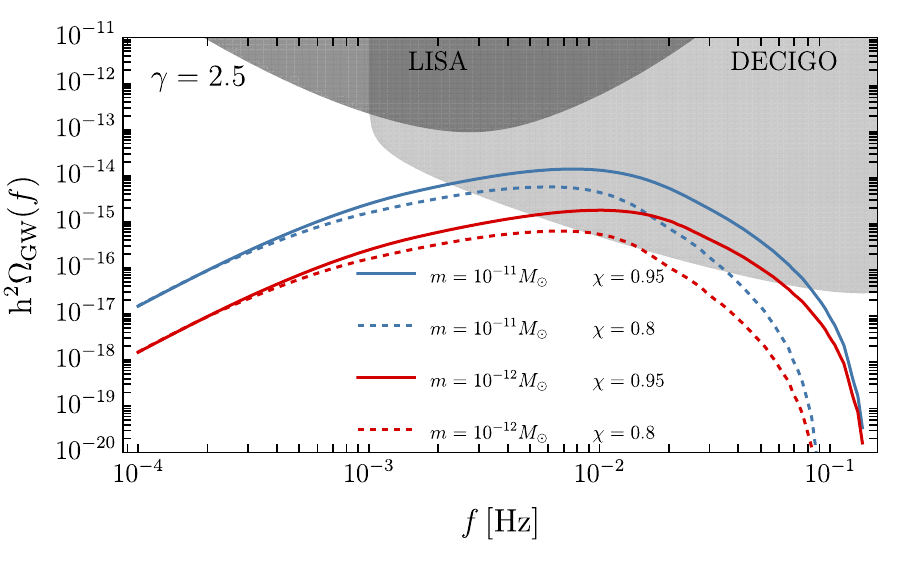}
		\caption{\textbf{Top panel:} GW background generated by PBHs of mass $m=10^{-11}\,M_\odot$ for the benchmark values of the spike slope $\gamma=2.5$ (red) and for $\gamma=2.75$ (blue), with SMBH spin $\chi=0.95$. The solid lines show the prediction, integrated over $5\times10^{5}\,M_\odot\leq M\leq10^{7}\,M_\odot$, and the shaded bands its theoretical uncertainty, $\Omega_{\rm GW}/4\leq\Omega_{\rm GW}^{\rm band}\leq6\,\Omega_{\rm GW}$.
        \textbf{Bottom panel:} Prediction for $\gamma=2.5$, for $m=10^{-11}\,M_\odot$ (blue) and $m=10^{-12}\,M_\odot$ (red), with $\chi=0.95$ (solid lines) and $\chi=0.8$ (dashed lines), integrated over $5\times10^{5}\,M_\odot\leq M\leq10^{7}\,M_\odot$. 
        In both panels we sum the multipoles up to $\ell=8$, retaining the modes $\tilde m=\pm\ell$, and use the optimistic mass function of Ref.~\cite{Caputo:2026rle}.
        The dark and light grey regions show the power-law integrated sensitivity curves of LISA~\cite{Babak:2021mhe,LISACosmologyWorkingGroup:2024hsc,LISACosmologyWorkingGroup:2025vdz} and DECIGO~\cite{Seto_2001,Kawamura:2020pcg}. For LISA we consider an ${\rm SNR}=10$ and an observation time of $4$ years while for DECIGO an ${\rm SNR}=1$ and $1$ year.}
        \label{fig:Omega_final}
	\end{center}
\end{figure}

In Fig.~\ref{fig:Omega_final} we show the GW background of Eq.~\eqref{eq:inclined_omega_gw} together with the power-law integrated sensitivities of LISA and DECIGO. All curves\footnote{We verified that multipoles up to $\ell = 6$ already model the spectrum well in the range of interest; we include up to $\ell = 8$ for a more accurate modelling of the UV tail.} include the multipoles up to $\ell=8$, with the modes $\tilde m=\pm\ell$. 
Our curves are obtained considering only the contribution coming from SMBH masses $5\times10^{5}\,M_\odot\leq M\leq10^{7}\,M_\odot$ with the optimistic mass function of Ref.~\cite{Caputo:2026rle}. 
For these masses X-ray reflection measurements suggest very high spins~\cite{Sisk-Reynes:2026rsq}, so we take $\chi=0.95$ as fiducial value. 
Such a high spin is less justified outside this mass range, and restricting the integration to it gives a conservative estimate of the signal. 
The implications of the mass cut and an estimate of the GW signal in the entire mass range are discussed in App.\,\ref{app:mass}. 

\subsection{Dependence on the physical parameters}
\paragraph{Spike slope.} The top panel of Fig.\,\ref{fig:Omega_final} shows how the signal depends on the slope of the spike. Our benchmark value is $\gamma=2.5$, the steepest profile produced by the adiabatic growth of the SMBH~\cite{Gondolo:1999ef}. 
We also show $\gamma=2.75$ as a phenomenological example of a steeper profile~\cite{Alexander:2008tq}. 
The signal grows quickly with $\gamma$, because a steeper profile places more PBHs at small radii, where the orbital frequency and the emitted power are higher. 
Going from $\gamma = 2.5$ to $\gamma = 2.75$ raises the peak of the spectrum by an order of magnitude. 
As a consequence, for $\gamma=2.75$ the central prediction enters the LISA sensitivity region, while for $\gamma=2.5$ it remains slightly below the sensitivity curve.
DECIGO, instead, can probe both configurations thanks to its lower sensitivity curve.
The shaded bands show the uncertainty on the normalisation of the signal. 
Three effects can increase it:
\begin{itemize}
\item Close to a rapidly rotating black hole, the density of the spike computed in the Kerr geometry exceeds the Schwarzschild one by a factor of two to three~\cite{Ferrer:2017xwm}.
\item The outer radius of the spike can be identified with the Bondi radius $r_{\rm B}=2GM/c_s^2$, or with $r_0=2GM/v_{\rm dm}^2=(5/9)\,r_{\rm B}$. 
The second choice is appropriate for a collisionless population, whose effective sound speed is $c_{s,\rm eff}^2=(5/9)\,v_{\rm dm}^2$ with $v_{\rm dm}^2=3\sigma_{\rm DM}^2$~\cite{Feng:2024obn, Lenoci:2023gjz}. 
Since $\rho_0r_0^\gamma\propto r_0^{\gamma-3}$, the smaller radius enhances the signal by $(5/9)^{\gamma-3}$.
\item We neglect the modes with $\tilde m\neq\pm\ell$ or $k\neq0$. 
In App.~\ref{app:polar_modes} we find that the polar modes $\ell=2$, $\tilde m=0$, $k=\pm2$ alone carry about one half of the power of the azimuthal modes $\ell=2$, $\tilde m=\pm2$, $k=0$. 
\end{itemize}
The product of these three points, in the fiducial scenario $\gamma=2.5$, is around $3\times1.3\times1.5\simeq6$.
In the opposite direction, the main uncertainty comes from the SMBH mass function. 
Replacing the mass function with the power-law fit~\cite{Babak:2017tow,Caputo:2026rle} 
$\dd n/\dd\ln M=0.005\,(M/4\times10^{6}\,M_\odot)^{-0.3}\,{\rm Mpc}^{-3}$
lowers the signal in the LISA band by about a factor of four. 
The band extends from $\Omega_{\rm GW}/4$ to $6\,\Omega_{\rm GW}$.

\paragraph{PBH mass and SMBH spin.} The bottom panel of Fig.\,\ref{fig:Omega_final} fixes $\gamma=2.5$ and compares $m=10^{-11}\,M_\odot$ and $m=10^{-12}\,M_\odot$, for our fiducial spin $\chi=0.95$ and for $\chi=0.8$, the weighted average of the spins reported in Ref.~\cite{Sisk-Reynes:2026rsq}. 

Close to the peak the growth with $m$ is slower, because heavier PBHs inspiral faster and have already emptied the innermost orbits, see Eq.~\eqref{eq:dNdr_evolved}. 
A larger spin brings the ISSO closer to the horizon, where the orbital frequency and the emitted power are higher. 
This raises the peak and extends the spectrum to higher frequencies. 
At low frequencies the signal is linear in $m$ and the emission comes from PBHs far from the SMBH. There the spin plays no role, and the curves for the two values of $\chi$ coincide. 
At the peak, lowering $\chi$ from $0.95$ to $0.8$ reduces the signal by approximately a factor of two. 
Within the range of parameters considered, the detectability of the signal depends more strongly on the spike slope rather than the SMBH spin. 
A detailed modeling of the spike slope is therefore essential to determine whether the GW background can be detected.

\subsection{Signal-to-noise ratio analysis for LISA}\label{sec:SNR}
To quantify the detectability with LISA we compute the signal-to-noise ratio (SNR) of the signal for a given slope and spin. The SNR is defined as
\begin{equation}
\label{Eq:SNR}
{\rm SNR}=\sqrt{T_{\rm obs}}\,\sqrt{\int\dd f\left(\frac{h^2\Omega_{\rm GW}(f)}{h^2\Omega_{\rm noise}(f)}\right)^2}\,.
\end{equation}
When foregrounds are absent the noise corresponds to the detector noise and thus we can write
\begin{equation}
h^2\Omega_{\rm noise}(f)=\frac{4\pi^2f^3}{3H_0^2/h^2}\,S_h(f)\,,
\end{equation}
and $S_h(f)$ is the LISA strain sensitivity~\cite{Babak:2021mhe,LISACosmologyWorkingGroup:2024hsc,LISACosmologyWorkingGroup:2025vdz}. Astrophysical foregrounds act as a further source of noise, reducing the sensitivity to stochastic signal. Different foregrounds are expected in the LISA band; in our estimates we include an extragalactic foreground from black-hole and neutron-star binaries~\cite{Chen:2018rzo,Babak:2023lro}, a Galactic foreground from double white-dwarf binaries~\cite{Evans:1987qa,Bender:1997hs,Karnesis:2021tsh}, and an extragalactic foreground from double white-dwarf binaries~\cite{Boileau:2025jkv}. 
To account for the presence of foregrounds we include their spectra to the LISA noise in $\Omega_{\rm noise}$ in Eq.~\eqref{Eq:SNR}, following the analytical shapes provided by~\cite{Pieroni:GWB_templates}.

\begin{figure*}[htb]
\begin{center}
\includegraphics[width=0.85\textwidth]{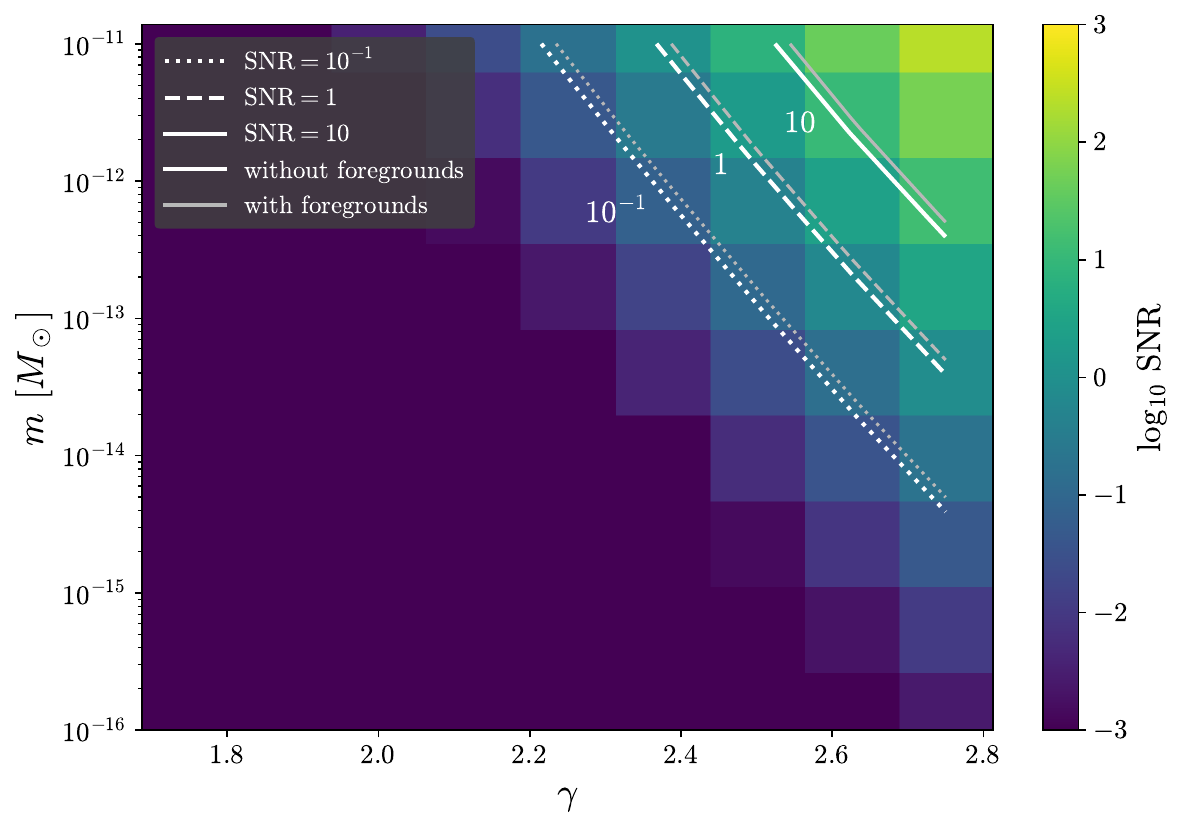}
    \caption{Signal-to-noise ratio of the GW background in LISA for $T_{\rm obs}=4\,$yr, as a function of the spike slope $\gamma$ and of the PBH mass $m$, up to the upper edge of the asteroidal window. The color map shows the SNR without astrophysical foregrounds. The grey dotted, dashed and solid lines are the contours ${\rm SNR}=0.1$, $1$ and $10$ including the foregrounds, and the thin white lines show the same contours without the inclusion of foregrounds. The SMBH spin is $\chi=0.95$ and the other parameters are as for the solid lines in the upper panel of Fig.~\ref{fig:Omega_final}.}
    \label{fig:SNR}
\end{center}
\end{figure*}

Figure~\ref{fig:SNR} shows the SNR for $T_{\rm obs}=4\,$yr as a function of the slope $\gamma$ and of the PBH mass, up to the upper edge of the asteroidal window, $m=10^{-11}\,M_\odot$. 
Since the signal is almost linear in $m$ where LISA is most sensitive, the SNR grows by about a factor of ten for each decade in $m$. 
The foregrounds lower the SNR by about $25\%$, so the contours with and without foregrounds almost overlap. 
For $\gamma\lesssim2.4$ the SNR remains below unity for every PBH mass in the window. 
Only sufficiently steep spikes place enough PBHs close to the central BH to produce a large SNR. 
There the orbital frequency is higher and the emitted power grows rapidly with decreasing radius, $L_{\rm GW}\propto r^{-5}$, so these PBHs dominate the signal. 
For the slope $\gamma=2.5$ an SNR above one requires $m\gtrsim2\times10^{-12}\,M_\odot$, and at $m=10^{-11}\,M_\odot$ the SNR is about five. 
An SNR of ten requires $\gamma\gtrsim2.55$ for $m=10^{-11}\,M_\odot$ and $\gamma\simeq2.7$ for $m=10^{-12}\,M_\odot$, and it is not reached for $m\lesssim5\times10^{-13}\,M_\odot$ within the slope range we consider.

\section{Conclusions}
\label{Sec::Conclusions}
In this work we have computed the stochastic GW background generated by asteroidal-mass PBHs that coalesce into the SMBH at the center of galaxies, after accumulating in the surrounding DM spikes. 
Although these systems have extremely small mass ratios and cannot be resolved individually, their combined emission over all galaxies produces a background in the millihertz band.

We have computed the GW background numerically for generic inclined spherical orbits in the Kerr geometry. 
We have also derived an analytical approximation based on equatorial prograde orbits, which have the smallest ISCO and give the largest contribution among all inclinations. 
This approximation reproduces the frequency dependence and peak amplitude of the numerical result.

The amplitude depends strongly on the spike slope. 
For our fiducial predictions, detection with LISA requires slopes above $\gamma=2.5$, the largest value produced by adiabatic SMBH growth. 
Even a modest increase above this value can raise the GW background enough to make it detectable. 
Nevertheless, these predictions remain subject to the theoretical uncertainties of the model. 
A more complete calculation, including all the emission modes, a proper integration over the full SMBH mass range (that should involve a mass distribution depending on spin and redshift) and a spike profile computed in the Kerr geometry, could bring the signal within reach of LISA also for the value $\gamma=2.5$. 
DECIGO can instead detect the signal for all the parameter configurations studied. 
In this case, a proper understanding of the spin distribution of low mass SMBHs would be needed, since it would affect the shape of the final spectrum.

This GW background provides a way to probe asteroidal-mass PBH dark matter independently of its formation mechanism and its detectability depends on the properties of the DM spikes around SMBHs, with the spike slope playing the main role in the scenarios considered.

\vskip 0.8cm
\centerline{\bf Acknowledgments}
\vskip 0.2cm
\noindent
A.R. and A.I. acknowledge support from the Swiss National Science Foundation (project number CRSII5\_213497). 
A.J.I.'s research is funded by Tamkeen under the NYU Abu Dhabi Research Institute grant ADHPG-AD457. G.P. is supported by the Estonian Research Council grants PSG869, KOHTO34, TARISTU24-TK3, TARISTU24-TK10, and the Center of Excellence program TK202.
D.P. is supported by the Swiss National Science Foundation under grants no. P500-2\texttt{\char`_}239221.
The Center of Gravity is a Center of Excellence funded by the Danish National Research Foundation under grant no. DNRF184.
\appendix
\begin{center}
 {\bf\Large Appendices}   
\end{center}
\section{Relativistic correction to the flux}
\label{app:flux}
\begin{figure*}[htbp]
\includegraphics[width=0.49\textwidth]{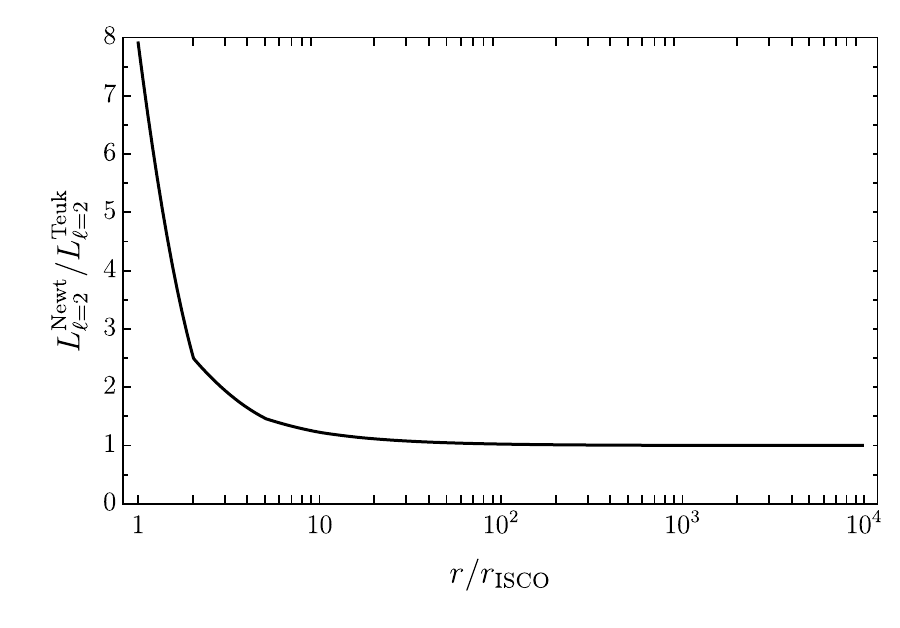}
\includegraphics[width=0.49\textwidth]{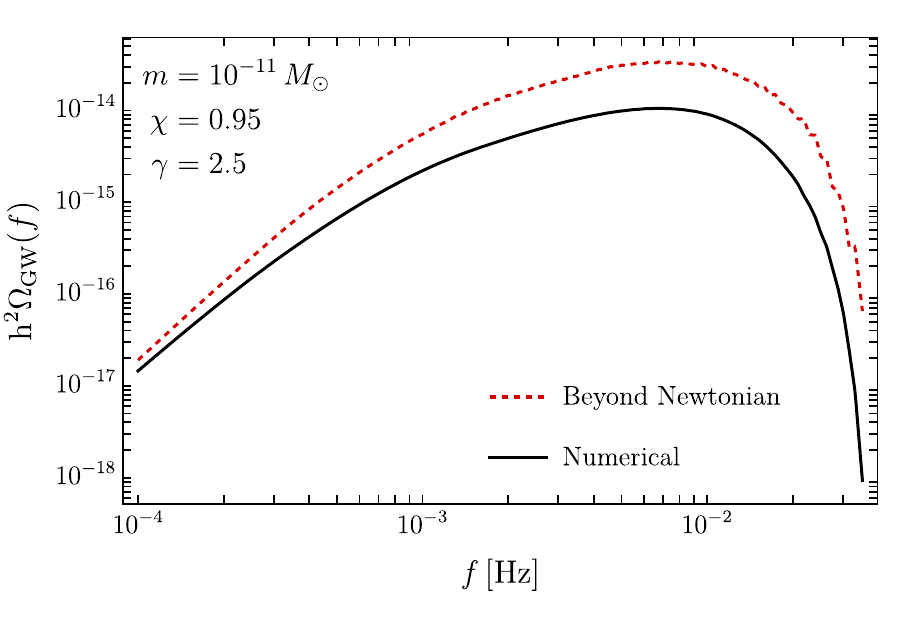}
    \caption{
    \textbf{Left panel:} Ratio between the Newtonian quadrupole luminosity and the numerical Teukolsky flux for the dominant modes $\ell=2$, $\tilde m=\pm2$ of an equatorial prograde orbit, as a function of the orbital radius. \textbf{Right panel:} Stochastic GW background generated by the PBH spike for $m=10^{-11}\,M_\odot$, $\chi=0.95$ and $\gamma=2.5$. The dashed red curve is the analytical result for equatorial prograde orbits, corrected with the radius-dependent luminosity ratio of the left panel, and the solid black curve is the numerical calculation of Sec.~\ref{sec:inclined_background}.}
    \label{fig:relativistic_correction}
\end{figure*}

In this appendix we quantify the breakdown of the Newtonian approximation in the strong-field regime and use the relativistic correction to assess the accuracy of the analytical prediction for the GW background. 
The left panel of Fig.~\ref{fig:relativistic_correction} shows the ratio between the Newtonian quadrupole luminosity and the numerical Teukolsky flux for an equatorial prograde orbit, $x=1$, considering the dominant modes $\ell=2$, $\tilde m=\pm2$. 
At large radii the ratio approaches unity, as expected in the weak-field limit, while towards the ISCO the relativistic effects become important and the Newtonian luminosity exceeds the Teukolsky flux by about a factor of eight.

In the right panel we compare the GW background obtained from the full numerical calculation of Eq.~\eqref{eq:inclined_omega_gw}, including all the orbital inclinations and using only the quadrupolar contributions $\ell=2$, $\tilde m=\pm2$, with the analytical approximation based only on equatorial prograde orbits, where now, contrary to Fig.~\ref{fig:Num}, the analytical luminosity is corrected with the radius-dependent ratio shown in the left panel. 
The corrected analytical result reproduces in a good approximation the amplitude at the peak of the numerical spectrum, within a factor of about three. 
This result shows that the prograde configuration captures the dominant part of the emission once the luminosity is computed in the strong-field regime. 
At the same time, the comparison shows that a relativistic flux is needed for reliable predictions close to the ISCO, where the Newtonian approximation breaks down.
The remaining discrepancy between the numerical and the analytical curve is expected, since the prograde approximation gives an upper estimate of the emission, whereas the numerical result averages over all orbital inclinations.
We notice also that at low frequencies, the two curves coincide as the Newtonian limit is completely recovered. 
\section{Estimate of the polar-mode contribution}
\label{app:polar_modes}

In the numerical calculation of the main text we retain the dominant azimuthal modes $\tilde m=\pm\ell$ with $k=0$. To estimate the contribution neglected by this truncation we compute separately the leading polar modes
\begin{equation}
(s,\ell,\tilde m,k,n)=(-2,2,0,\pm2,0),
\end{equation}
which are relevant especially for polar orbits, $x\simeq0$.
For spherical orbits $\Omega_r=0$, and for $\tilde m=0$ the frequency of the mode is entirely determined by the polar motion. Restoring the dimensions, it reads
\begin{equation}
f_{k}(r,x)
=
\frac{c^3}{2\pi GM}
\left|k\,\Omega_\theta(r,x)\right|.
\label{eq:polar_frequency}
\end{equation}
The two modes $k=\pm2$ have the same physical frequency once the absolute value is taken, and their fluxes add up to the total luminosity at infinity
\begin{equation}
L_{ k}^{\infty}(r,x)
=
\frac{c^5}{G}
\left(\frac{m}{M}\right)^2
\left[
\mathcal F^\infty_{2,0,2,0}(r,x)
+
\mathcal F^\infty_{2,0,-2,0}(r,x)
\right].
\label{eq:polar_luminosity}
\end{equation}
At fixed inclination Eq.~\eqref{eq:polar_frequency} can be inverted numerically to obtain $r_{k}=r(f_{k},x)$, and the inclination-integrated spectral luminosity of the polar modes is
\begin{align}
\mathcal L_{2,0,\pm2}(f,M)
={}&
\int_{-1}^{1}\dd x\;
L_{k}^{\infty}
\left[r_{k}(f,x),x\right]
\left.
\frac{\dd^2N}{\dd r\,\dd x}
\right|_{r_{k}(f,x)}
\left|
\frac{\partial r_{k}}{\partial f}
\right|_x
\nonumber\\
&\times
\Theta\!\left(f-f_{\rm min}(x)\right)
\Theta\!\left(f_{\rm ISSO}(x)-f\right).
\label{eq:polar_spectral_luminosity}
\end{align}
We compare this result with the inclination-integrated luminosity of the dominant quadrupolar azimuthal modes, $(\ell,\tilde m,k)=(2,\pm2,0)$, and at each frequency in their common domain we find
\begin{equation}
\mathcal R
=
\frac{\mathcal L_{2,0,\pm2}(f,M)}{\mathcal L_{2,\pm2,0}(f,M)
}\simeq 0.5.
\label{eq:polar_azimuthal_ratio}
\end{equation}
The polar modes alone contribute about one half of the dominant azimuthal component, which shows that the modes omitted from the calculation can increase the overall normalisation of the GW background and motivates their inclusion in the upper edge of the uncertainty band.
We therefore conclude that the dominant energetic contribution comes from the quadrupolar azimuthal modes $\ell=2$, $\tilde m=\pm2$ for equatorial prograde orbits, $x=1$.
\section{The spin and the full mass range}\label{app:mass}
In the main text we integrated Eq.\,(\ref{eq:inclined_omega_gw}) only for SMBH masses in the range $5\times10^{5}\,M_\odot\leq M\leq10^{7}\,M_\odot$. For these masses X-ray reflection measurements indicate high spins~\cite{Sisk-Reynes:2026rsq} that we have fixed correspondingly to $\chi=0.95$. Outside this range the spin is poorly known, even though there is a tendency towards lower spins for lower masses. 
Since the spin fixes the position of the ISSO and is crucial to evaluate the resulting GW spectrum, in this appendix we estimate the GW signal in the entire mass range.

To do so, we extend the integration of Eq.\,\ref{eq:inclined_omega_gw} to the full range $10^{4}\,M_\odot\leq M\leq10^{9}\,M_\odot$, considering $\chi=0.95$ and $\chi = 0.5$. 
For both missing ranges, i.e. for SMBHs of masses $10^{4}\,M_\odot\leq M\leq5\times10^{5}\,M_\odot$ and of masses $10^{7}\,M_\odot\leq M\leq10^{9}\,M_\odot$, we estimate the GW spectrum produced for two different values of the spin, based on~\cite{Sisk-Reynes:2026rsq}: a maximal one $\chi=0.95$ and a lower one $\chi=0.5$. The spectra evaluated at these two values determine the band in which the effective spectrum is expected. In this way, not only we estimate the impact of those masses on the final spectrum, but also the corresponding frequency region which would be more affected.

\begin{figure}[thb!]
	\begin{center}
\includegraphics[width=.99\textwidth]{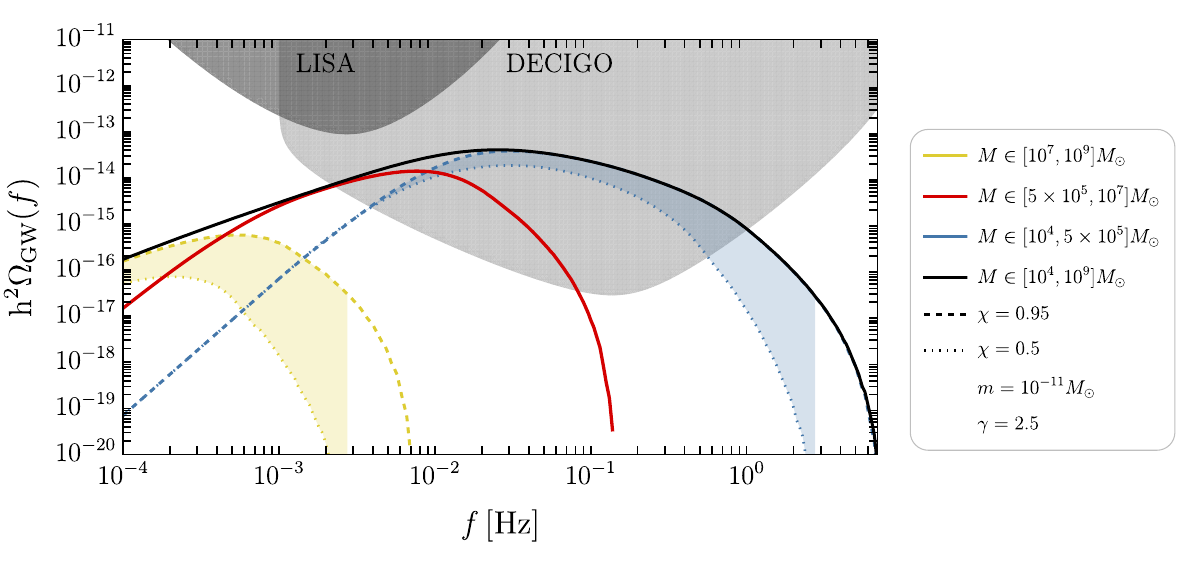}
		\caption{GW background generated by PBHs of mass $m=10^{-11}\,M_\odot$ for a benchmark value of the spike slope $\gamma=2.5$. The red curve corresponds to our fiducial spectrum, integrated over $5\times10^{5}\,M_\odot\leq M\leq10^{7}\,M_\odot$ with $\chi=0.95$. The yellow and blue bands show the contributions of SMBHs with $10^{7}\,M_\odot\leq M\leq10^{9}\,M_\odot$ and $10^{4}\,M_\odot\leq M\leq5\times10^{5}\,M_\odot$, respectively. Their upper (dashed) and lower (dotted) edges correspond to $\chi=0.95$ and $\chi=0.5$. The black curve shows the total signal for $10^{4}\,M_\odot\leq M\leq10^{9}\,M_\odot$ and $\chi=0.95$.The dark and light grey regions show the power-law integrated sensitivities of LISA~\cite{Babak:2021mhe,LISACosmologyWorkingGroup:2024hsc,LISACosmologyWorkingGroup:2025vdz} and DECIGO~\cite{Seto_2001,Kawamura:2020pcg}, respectively.} 
        \label{fig:Omega_final1}
	\end{center}
\end{figure}

The results are shown in Fig.~\ref{fig:Omega_final1}. In red we show our benchmark signal, obtained for $5\times10^{5}\,M_\odot\leq M\leq10^{7}\,M_\odot$ and $\chi=0.95$. The band corresponding to higher masses is reported in yellow. 
These masses would mostly contribute to the IR tail, with a quite negligible contribution in the range of interest. The band for lower masses, instead, is reported in blue. Those would dominate the signal at higher frequencies and have a major effect observable by DECIGO. 
We checked that, even in the most optimistic case, these contributions would leave the portion of the spectrum that falls within the LISA sensitivity band practically unchanged. 
Neglecting them therefore does not affect the analysis we report in Sec.\,\ref{sec:SNR}. The situation is different for DECIGO, whose sensitivity covers the frequencies where the light SMBHs dominate. 
There the band gives the expected range of the signal, while a model of the spin distribution of light SMBHs would be needed to predict its detailed shape. For this reason in Sec.\,\ref{sec:SNR} we limited our analysis to the LISA case. 
For reference, we also report in black the total signal that would be observed if the spin of all the SMBH in the whole mass range would follow a monochromatic distribution with $\chi=0.95$.

\newpage
\bibliographystyle{JHEP}
\bibliography{references}
\end{document}